\documentclass{emulateapj}
\RequirePackage{lineno}
\usepackage{amssymb}
\usepackage{amsmath}
\usepackage{mathrsfs}
\usepackage{natbib}
\usepackage[colorlinks, linkcolor=blue, urlcolor=blue, citecolor=blue]{hyperref}
\usepackage{array}
\usepackage{float}
\usepackage{graphicx}
\usepackage{subfigure}
\usepackage{rotating}
\usepackage{color}
\usepackage[all]{hypcap}
\usepackage[toc,page]{appendix}

\begin{document}

\title{Photospheric Radius Evolution of Homologous Explosions }
\author{Liang-Duan Liu\altaffilmark{1,2,3}, Bing Zhang\altaffilmark{3}, Ling-Jun Wang\altaffilmark{4}, and Zi-Gao Dai\altaffilmark{1,2}}

\begin{abstract}
Recent wide-field surveys discovered new types of peculiar optical transients that showed diverse behaviors of the evolution of photospheric properties. We develop a general theory of homologous explosions with constant opacity, paying special attention on the evolution of the photospheric radius $R_{\rm ph}$. We find that regardless of the density distribution profile, $R_{\rm ph}$ always increases early on and decreases at late times.
This result does not depend on the radiation and cooling processes inside the ejecta.The general rising/falling behavior of $R_{\rm ph}$ can be used to quickly diagnose whether the source originates from a supernova-like explosion. The shape of the $R_{\rm ph}$ evolution curve depends on the density profile, so the observations may be used to directly diagnose the density profile as well as the temperature profile of the ejecta. All the well-monitored supernovae show such a $R_{\rm ph}$ rising/falling behavior, which is consistent with our theory. The recently discovered peculiar transient AT2018cow showed a continuous decay of $R_{\rm ph}$, which is disfavored to be of a supernova-like explosion origin. Our result therefore supports the interpretation of this transient as a tidal disruption event.
\end{abstract}


\affil{\altaffilmark{1}School of Astronomy and Space Science,
Nanjing University, Nanjing 210093, China;
dzg@nju.edu.cn}
\affil{\altaffilmark{2}Key Laboratory of Modern Astronomy and
Astrophysics (Nanjing University), Ministry of Education, China}
\affil{\altaffilmark{3}Department of Physics and Astronomy,
University of Nevada, Las Vegas, NV 89154, USA; zhang@physics.unlv.edu}
\affil{\altaffilmark{4}Astroparticle Physics,
Institute of High Energy Physics,
Chinese Academy of Sciences, Beijing 100049, China}
%\affil{\altaffilmark{5}Guangxi Key Laboratory for Relativistic Astrophysics, Department of Physics, Guangxi University, Nanning 530004, China}
%\affil{\altaffilmark{6} Institute of Astrophysics, Central China Normal University, Wuhan 430079, China}

\section{Introduction}

\label{sec:Intro}

The rapid development of several wide-field optical surveys (e.g.  the intermediate Palomar Transient Factory (iPTF)\footnote{\url{https://www.ptf.caltech.edu/iptf}.}, the All-Sky Automated Survey for Supernovae (ASAS-SN)\footnote{\url{http://www.astronomy.ohio-state.edu/~assassin/index.shtml}.},  the Panoramic Survey Telescope \& Rapid Response System (Pan-STARRS)\footnote{\url{https://panstarrs.stsci.edu}.}, and Dark Energy Survey (DES)\footnote{\url{https://www.darkenergysurvey.org}.})
%\footnote{ Some well known transient survey projects as follows:
%\begin{itemize}
  %\item The intermediate Palomar Transient Factory (iPTF)
  %\url{https://www.ptf.caltech.edu/iptf}
  %\item The All-Sky Automated Survey for Supernovae (ASAS-SN)
  %\url{http://www.astronomy.ohio-state.edu/~assassin/index.shtml}
  %\item The Panoramic Survey Telescope \& Rapid Response System (Pan-STARRS) \url{https://panstarrs.stsci.edu}
  %\item Dark Energy Survey (DES)
  %\url{https://www.darkenergysurvey.org}
%\end{itemize}}
is revolutionizing the field of time-domain transient astrophysics. Besides known objects (e.g. supernovae (SNe) and tidal disruption events (TDEs)) with extreme properties (e.g. ASASSN-15lh, \citealt{Dong2016} and  iPTF14hls, \citealt{Arc2017}) these observations have also discovered several peculiar, rapidly evolving, luminous transients whose nature is not properly understood \citep{Dro2014,Arc2016,Whi2017}. One example is
%timescales and luminosities are difficult to determine  the nature of them
AT2018cow, which showed a very rapid rise of the lightcurve, and a steady decay of the photospheric radius $R_{\rm ph}$ \citep{Pre2018,Per2018,Kui2018}. Such a behavior has never been observed before in a supernova. Possible interpretations range from special types of explosions to special types of TDEs, but no definite conclusion has been drawn.

%Some basic physical parameters of  the supernova (SN) explosions can be obtained by comparing the observations with theoretical computations of light curves and/or spectra.

Here we develop a simple theory of the evolution of the photospheric radius, $R_{\rm ph}$, of a generic explosion, which is homologous (each layer expanding with a constant velocity) but could have arbitrary density profile, heating/cooling structure, and hence,  arbitrary temperature profile. Assuming a constant opacity, we derive a generic behavior of the $R_{\rm ph}$ evolution of such explosions. Section \ref{sec:mod} presents the general theory. Section \ref{sec:res} presents several specific density profile examples. The results are summarized in Section \ref{sec:dis} with some discussions on its application to AT2018cow and other transients.

%In Section \ref{sec:mod}, we present the general behavior of the photospheric radius evolution with an arbitrary density profile. Section \ref{sec:res} shows the photospheric radius evolution in the specific density profiles. The conclusions and discussions are given in Section \ref{sec:dis}.

\section{A general theory of photospheric radius evolution}
\label{sec:mod}

Observationally, the photospheric radius at a particular time can be derived by $R_{\rm ph}(t)=[L_{\rm bol}(t)/4 \pi \sigma T_{\rm eff}^{4}(t)]^{1/2}$, where the bolometric luminosity $L_{\rm bol}(t)$ can be derived from the multi-color photometry at each epoch $t$, and the effective temperature $T_{\rm eff}(t)$ can be inferred by fitting the spectra at the same epoch. From the theoretical model, the photospheric radius evolution depends on the dynamical evolution and the density profile of the ejecta but is independent of the cooling and heating processes\footnote{This statement is strictly correct for photospheric emission. In practice, however, $R_{\rm ph}$ is determined by the ``observed'' $L_{\rm bol}$ and $T_{\rm eff}$, which may be dominated by the contribution of emission outside the photospheric radius (i.e. the ejecta layers already in the so-called ``nebula'' phase) during the late phase of a supernova explosion. In such cases, the effective $R_{\rm ph}$ derived from the data does depend on the heating process in the ejecta. }. In the literature \citep{Arn1982}, the photospheric radius is often described as
\begin{equation}
\label{Eq:Rph_0}
R_{\rm ph}(t)=R(t)-\frac{2}{3} \lambda(t),
\end{equation}
for an ejecta with a uniform, time-dependent density $\rho(t)$, where $\lambda(t)=1/\rho(t) \kappa$ is the mean free path of the photons, and $\kappa$ is the opacity. In reality, the density profile of an explosion is not uniform. Different types of density profiles will modify Eq.(\ref{Eq:Rph_0}) to much more complicated forms.
%Note that this equation is only valid for an uniform density profile of the ejecta. The  different density profiles may  have strongly influence on the photospheric radius evolution.

In order to simplify the problem to a tractable form, we make several assumptions in the following. First, the supernova ejecta is homologously expanding and spherically symmetric. Second, Thomson scattering dominates the opacity so that the opacity $\kappa$ is a constant throughout the evolution. Third, we assume that the emission from the ejecta layers above the photosphere in the nebula phase does not outshine the emission from the photosphere itself. Introducing more complicated scenarios would introduce more qualitative differences (see discussion in Section \ref{sec:dis}), but the general features discussed in this paper may not alter substantially.

For an energetic explosion such as a supernova, the ejecta would enter a homologous expansion phase after a few times of the expansion timescale $R_{\rm p}/v$, where $R_{\rm p}$ is the radius of the progenitor and $v$ is the mean expansion velocity of the ejecta. For a homologous expansion of the ejecta with a velocity gradient, fast ejecta layers propagate in front and slow ejecta layers lag behind. The inner boundary of the ejecta is defined by the slowest ejecta, and its radius reads
\begin{equation}
R_{\rm min} \left( t\right) =R_{\rm min,0} +v_{\text{min}}t,
\end{equation}
where $v_{\rm min}$ is the minimum velocity of the ejecta and $R_{\rm min,0}$ is the initial radius of the innermost radius when the explosion enters the homologous phase. The outer boundary of the ejecta is defined by
\begin{equation}
R\left( t\right) =R_0 +v_{\text{max}}t,
\end{equation}%
where $R_0 $ is the initial radius of the outermost radius in the homologous phase, and $v_{\text{max}}$ is the maximum velocity of the ejecta.
% outermost layer of the ejecta at any epoch $t$. And due to
The homologous expansion conditions imply $R_p < R_{\rm min,0} < R_0$ and $v_{\text{min}} \ll v_{\text{max}}$.

We define a comoving, dimensionless radius $x$ as%
\begin{equation}
x\equiv \frac{r-R_{\rm min}}{R-R_{\rm min}},
\end{equation}%
where $r$ is the radius of a particular layer in the ejecta from the center of explosion,  and $0\leq x\leq 1$ is satisfied for all the elements within the ejecta.

For a homologous expansion,  the density of the ejecta can be written as %\citep{Arn1980,Arn1982}
\begin{equation}
\rho \left( r,t\right) =\rho \left( R_0,0\right) \eta \left( x\right) \left[
\frac{R_0-R_{\min,0}}{R\left( t\right) -R_{\min}(t)}\right] ^{3},  \label{Eq:density}
\end{equation}%
where $\rho \left( R_0,0\right)$ is the initial density at the outermost radius of the ejecta,
$\eta \left( x\right) $ is a function to describe the density profile of the ejecta, and
for a uniform density distribution, one has $\eta \left( x\right) =1$.  The $\left[
(R_0 - R_{\min,0}) /(R\left( t\right) - R_{\rm min}(t)) \right] ^{3}$ scaling describes the homologous
expansion of the ejecta.

The total ejecta mass can be derived through integrating over the density profile, i.e.
\begin{eqnarray}
M_{\text{ej}} &=&\int_{R_{\rm min}(t)}^{R\left( t\right) }4\pi r^{2}\rho \left(
r,t\right) dr  \notag \\
&=&\left[ 4\pi \rho \left(R_0,0\right) R_0^{3} \right] I_{\text{M%
}},  \label{Eq: mass}
\end{eqnarray}%
where
\begin{equation}
I_{\text{M}}\equiv \int_{0}^{1}x^{2}\eta \left( x\right) dx,
\end{equation}%
is a dimensionless factor for ejecta mass that is related to the assumed density profile.

The total kinetic energy  with a given density profile can be derived as
\begin{eqnarray}
E_{\text{K}}\left( t\right) &=&\int_{R_{\rm min}(t)}^{R\left( t\right) }\frac{1}{2}\rho
v^{2}4\pi r^{2}dr  \notag \\
&=&\left[ 2\pi \rho \left(R_0,0\right) R_0^{3}  \right] v_{\text{%
max}}^{2}I_{\text{K}},  \label{Eq: Ek}
\end{eqnarray}%
where
\begin{equation}
I_{\text{K}}\equiv \int_{0}^{1} x^4 \eta \left( x\right) dx,
\end{equation}%
is a dimensionless factor for kinetic energy that is related to the assumed density profile.
%Note that the first factor in the bracket of Eq.$\left( \ref{Eq: Ek}\right) $ has the dimension of mass.
%, and the second factor is square of the expansion velocity.

Combining Eq.$\left( \ref{Eq: mass}\right) $ and Eq.$\left( \ref{Eq: Ek}%
\right)$, the velocity of the outermost layer of the ejecta (which is the maximum velocity in the ejecta) is given by%
\begin{equation}
v_{\text{max}}=\left( \frac{2E_{\text{K}}}{M_{\text{ej}}}\frac{I_{\text{M}}}{%
I_{\text{K}}}\right) ^{1/2}.
\end{equation}%
For a uniform density distribution, one has $v_{\text{max}}=\left( 10E_{\text{K}}/3M_{\text{ej}%
}\right) ^{1/2}.$ It is worth noting that  $v_{\text{max}}$ is a parameter in our
semi-analytic model, which usually cannot be measured directly.

The total optical depth of the ejecta $\tau _{\text{tot}}$ is%
\begin{equation}
\tau _{\text{tot}}=\int_{R_{\rm min}(t)}^{R\left( t\right) }\kappa \rho dr.
\end{equation}%
For a constant opacity $\kappa$, one has
%and a dimensionless density profile $\eta \left( x\right) ,$ we have
\begin{equation}
\tau _{\text{tot}}\left( t\right) =\tau _{\text{tot}}\left( 0\right) \left[
\frac{R_0-R_{\min,0} }{R\left( t\right) -R_{\min}(t)}\right] ^{2}, \label{Eq:tau_tot}
\end{equation}%
where $\tau _{\text{tot}}\left( 0\right) $ is the initial optical depth, i.e.,
\begin{equation}
\tau _{\text{tot}}\left( 0\right) =\kappa \rho \left(R_0,0\right) R_0 I_{\tau },
\end{equation}%
and%
\begin{equation*}
I_{\tau }\equiv \int_{0}^{1}\eta \left( x\right) dx
\end{equation*}%
is a dimensionless factor for the optical depth that is related to the density profile.  The total optical
depth $\tau _{\text{tot}}\left( t\right) $ decreases with time following $%
t^{-2}$.  When $\tau _{\text{tot}}=2/3$ the whole ejecta becomes
transparent. We introduce a critical time so that $\tau _{\text{tot}}(t_{\tau
})=2/3$ is satisfied, which reads %
\begin{equation}
t_{\tau }= \left( \frac{R_0-R_{\min,0} }{v_{
\text{max}}-v_{\min}} \right)\left \{ \left[ \frac{3\tau _{\text{tot}}\left( 0\right)  }{2}\right] ^{1/2}-1 \right \},
\end{equation}%
after $t_{\tau }$ the explosion enters the so-called ``nebular'' phase, when the assumption of blackbody emission becomes invalid.

Based on the Eddington approximation, the relation between the externally
observed effective temperature $T_{\text{eff}}$ and the internal temperature $T$ at an optical depth $\tau =2/3$
is given by \citep{Arn1980,Arn1989}
\begin{equation}
T^{4}=\frac{3}{4}T_{\text{eff}}^{4}\left( \tau +\frac{2}{3}\right).
\end{equation}%
%One can see that the observed temperature $T_{\text{eff}}$ is a good measure of the true
%temperature $T$ of ejecta.
Therefore, the location of the photospheric radius $R_{\text{ph}}$ is at $\tau \left( R_{\text{%
ph}}\right) =2/3,$ which is defined as
\begin{equation}
\int_{R_{\text{ph}}\left( t\right) }^{R\left( t\right) }\kappa \rho dr=\frac{%
2}{3}.
\end{equation}

Using Eq.(\ref{Eq:density}) and Eq.(\ref{Eq:tau_tot}), this condition can be rewritten as
\begin{equation}
\frac{\tau _{\text{tot}}\left( t\right) }{I_{\tau }}I_{\text{ph}}\left(
t\right) =\frac{2}{3},  \label{Eq:xph-dete}
\end{equation}%
where
\begin{equation}
I_{\text{ph}}\left( t\right) \equiv \int_{x_{\text{ph}}\left( t\right)
}^{1}\eta \left( x\right) dx,
\end{equation}%
and $x_{\text{ph}} = (R_{\rm ph}-R_{\rm min}) / (R-R_{\rm min})$ is a dimensionless parameter of the photosphere radius.
%the ratio of  the photospheric radius to the
%outermost radius of the ejecta in the comoving coordinate.
As the expansion proceeds, \ $x_{\text{ph }}$decreases with time, which means that the
photospheric radius recedes in the comoving coordinate of the ejecta.
When $t=t_{\tau }$, one has $x_{\text{ph}}=0,$ i.e. the photospheric radius
reaches the innermost radius of the ejecta, and the photons produced anywhere in the ejecta can escape directly without being scattered.

The photospheric radius is
\begin{equation}
R_{\text{ph}}\left( t\right) =[R\left( t\right)-R_{\rm min}(t)] x_{\text{ph}}\left( t\right)+R_{\rm min}(t).
\label{Eq:Rph}
\end{equation}%
The evolution of the photospheric radius depends on the competition between
the expansion and the recession of$\ x_{\text{ph}}$ in the comoving
coordinate of the ejecta.

The time derivative of the photospheric radius reads
\begin{eqnarray}
\frac{dR_{\text{ph}}}{dt} &=&\left( v_{\max }-v_{\min }\right) x_{\text{ph}}
\notag \\
&&+\left[ R\left( t\right) -R_{\min }\left( t\right) \right] \frac{dx_{\text{%
ph}}}{dt}+v_{\min }.
\label{Eq:vph}
\end{eqnarray}

It is worth noting that $dR_{\rm ph}/dt$ is not the so-called photospheric velocity $v_{\rm ph}$ as measured by observers based on spectral information, which is the instantaneous velocity of the layer of ejecta that reaches the photosphere radius. In our calculation, we have assumed that the ejecta is homologously expanding, which means the local velocity $v$ is proportional to the radius $r$.  Therefore, the photospheric velocity $v_{\rm ph}$ is given by 
\begin{equation}
\frac{v_{\rm ph}}{v_{\rm max}}= \frac{R_{\rm ph}-R_{\rm min}}{R-R_{\rm min}}.
\end{equation}

Comparing $v_{\rm ph}$ with the observational photospheric velocity evolution obtained from absorption spectral features could help us to constrain the velocity profile of the  explosion ejecta. 

Taking the time derivative of Eq.$\left( \ref{Eq:xph-dete}\right) $, one has
\begin{equation}
\frac{dx_{\text{ph}}}{dt}\frac{d}{dx_{\text{ph}}}\left[ \int_{x_{\text{ph}%
}}^{1}\eta \left( x\right) dx\right] =\frac{d}{dt}\left( \frac{2}{3}\frac{%
I_{\tau }}{\tau _{\text{tot}}}\right) .
\end{equation}

We can then obtain the time derivative of $x_{\text{ph}}$ as
\begin{equation}
\frac{dx_{\text{ph}}}{dt}=-\frac{4I_{\tau }}{3}\frac{v_{\text{max}}-v_{\min}}{R\left(
t\right)-R_{\rm min}(t) }\frac{1}{\eta \left( x_{\text{ph}}\right) \tau _{\text{tot}}\left(
t\right) }.
\end{equation}
Substituting  it into Eq.$\left( \ref{Eq:vph}\right) ,$ we get
\begin{equation}
\frac{dR_{\text{ph}}}{dt} =\left( v_{\text{max}}-v_{\min} \right)\left [ x_{\text{ph}}-\frac{4I_{\tau }}{3\eta
\left( x_{\text{ph}}\right) \tau _{\text{tot}}\left( t\right) }\right ] .
\label{v_ph}
\end{equation}

The location of the maximum photospheric radius is found by setting $dR_{\text{ph}}/dt  = 0$ in
Eq.$\left( \ref{v_ph}\right)$, giving
\begin{equation}
x_{\text{ph}}\left( t\right) -\frac{4I_{\tau }}{3\eta \left( x_{\text{ph}%
}\right) \tau _{\text{tot}}\left( t\right) }=0  \label{Eq:tau_tr1}
\end{equation}

Let us define a ``transitional'' optical depth $\tau _{\text{tr}}$ by $dR_{\text{ph}}/dt=0
$, i.e., when the total optical depth $\tau _{\text{tot}}$ equals $\tau _{%
\text{tr}},$ the photospheric radius reaches its maximum. Using Eq. $\left( %
\ref{Eq:tau_tr1}\right)$, we have
\begin{equation}
\tau _{\text{tr}}=\frac{4I_{\tau }}{3\eta \left( x_{\text{ph}}\right) x_{%
\text{ph}}\left( t\right) }.  \label{Eq:tau_tr2}
\end{equation}
We can then find out the time $t_{\text{ph,max}}$  when the
photospheric radius reaches the maximum by substitution Eq.(\ref{Eq:tau_tr2})
into Eq.(\ref{Eq:tau_tot}).

Therefore, according to our general theory, we reach the following conclusion:
\textit{In an ejecta undergoing homologous expansion, for an arbitrary density distribution
profile, the photospheric radius $R_{\rm ph}$ always displays an initially rising phase and a later declining phase.}
The result does not depend on the radiation and cooling process inside the ejecta.

\section{Examples}
%The photospheric radius evolution in the specific density profiles}
\label{sec:res}

%In above section, we have proofed that for an arbitrary density profile, the evolution of the photospheric radius has a similar evolution behavior (first rise and then fall).
In this section, we consider the $R_{\rm ph}$ evolution in several examples with different specific density profiles (a spherical symmetry is assumed throughout), and show the differences among these examples.
%caused by various density profiles in the ejecta.

\subsection{CASE I:  uniform density profile }

If the density profile of the ejecta is uniform, one has $\eta (x)=1$.  Substituting it into Eq.(\ref{Eq:xph-dete}), one can obtain $x_{\rm ph}=1-2/3\tau_{\rm tot}$. This is equivalent to Eq.(\ref{Eq:Rph_0}).  As the ejecta expands homologously, $R(t)$ linearly increases with time, while the mean free path of the photons evolves as $\lambda \propto t^3$. According to Eq.(\ref{Eq:tau_tr2}), we find that $\tau_{\rm tr}=2$ corresponding to the maximum photospheric radius. The evolution of the photospheric radius and velocity with an uniform density  is shown in Fig. \ref{fig:ETA} red dashed lines.

To calculate $R_{\rm ph}$, we need to solve Eq.(\ref{Eq:xph-dete}) and then apply Eq. (\ref{Eq:Rph}). For $v_{\rm min} \ll v_{\rm max}$, given a certain density profile $\eta(x)$, there are four main free parameters that may significantly affect the results:  the ejecta mass $M_{\rm ej}$, the initial radius of the outer layer of the ejecta $R_0$, the initial kinetic energy $E_{\rm K}$, and the opacity $\kappa$.
%Assuming that Thomson scattering dominates the opacity, the opacity $\kappa$ depends on the composition of the ejecta. For pure hydrogen gas as $\kappa \approx 0.4$ cm$^{2}$ g$^{-1}$, and for a He-dominated ejecta  $\kappa \approx 0.2$ cm$^{2}$ g$^{-1}$. In the following calculation  we use a fixed value of the opacity as $\kappa \approx 0.1$ cm$^{2}$ g$^{-1}$.
In the following, we investigate how different parameters affect the result for the uniform density case.
%For an uniform density profile, we test the effects of varying the free parameters.  The parameters were changed one by one using three different values while holding the others parameters constant.  Here
The fiducial parameters are chosen as (plotted with red dashed line in Fig \ref{fig:MRE}): $M_{\rm ej}=3.0M_{\odot}$; $E_{\rm K}=2\times 10^{51}$ erg; $R_0=10^{13}$cm; and $\kappa = 0.1$ cm$^{2}$ g$^{-1}$.

We first investigate the effect of ejecta mass. Three values are adopted: $M_{\rm ej} = 1, 3, 8 M_{\odot}$. The results are shown in panel (a) of Fig \ref{fig:MRE}. One can see that the ejecta mass has significant influence on the photospheric radius evolution. As $M_{\rm ej}$ increases, the maximum $R_{\rm ph}$ is larger and the time it takes to reach the maximum longer.
%As a result of  increasing the ejecta mass, the peaks of the photospheric radii become larger and the time of the ejecta becomes transparent become longer.

Next, we investigate the effect of kinetic energy by adopting $E_{\rm K}= 5 \times 10^{50},  2 \times 10^{51}, 4 \times 10^{51}$ erg. As shown in panel (b) of Fig \ref{fig:MRE},  $E_{\rm K}$ mainly influences the time when $R_{\rm ph}$ reaches the maximum, but
% peak time of the photospheric radius, and it
has little influence on the peak value of $R_{\rm ph}$. Because we fixed the ejecta mass as  $M_{\rm ej}=3.0M_{\odot}$, a higher kinetic energy corresponds to a larger velocity scale, resulting in a faster evolution of $R_{\rm ph}$.

The panel (c) of Fig \ref{fig:MRE} shows that initial radius $R_0$ has a negligible effect on the evolution of $R_{\rm ph}$. We adopt three values, i.e. $R_0 = 10^{12}, 10^{13}, 10^{14}$ cm, and find that $R_{\rm ph}$ essentially does not change. This is because during the evolution, we are mostly investigating the epochs when $vt \gg R_0$, so that the initial conditions do not matter much.

Finally, we consider the effect of opacity.  In the panel (d) of Fig \ref{fig:MRE}, three values of opacity is chosen as $\kappa =  0.1, 0.2,  0.4$ cm$^{2}$ g$^{-1}$. We can see that a higher opacity leads to a higher maximum $R_{\rm ph}$ and a longer time to reach it.

In all these cases, the shape of the $R_{\rm ph}$ evolution curves remain the same, which only depends on the density profile function $\eta(x)$.

\subsection{CASE II:  broken power law density profile}

We next relax the assumption of uniform density distribution. The first case we study is
%that the ejecta of mass $M_{\rm ej}$ is spherical symmetric and the density profile is
a broken power law density profile, with a flatter profile in the inner
region and a steeper profile in the outer part of the ejecta \citep[e.g.][]{Che1982,Mat1999,Kas2010,Mor2013},
\begin{equation}
\eta \left( x\right) =\left\{
\begin{array}{cc}
\left( x/x_{0}\right) ^{-\delta } & 0\leq x\leq x_{0}, \\
\left( x/x_{0}\right) ^{-n} & x_{0}\leq x\leq 1,%
\end{array}%
\right.
\end{equation}%
where $x_{0}$ is the dimensionless transition radius from the inner region
to the outer region. %to $R\left( t\right)$.
Only for $n>5$ and $\delta <3$ the conditions of finite energy  and mass can be satisfied.  Such a profile is often adopted in modeling SNe. The outer density index $n$
depends on the progenitor of the SN.  For SN Ib/Ic and SN Ia progenitors, one has $n\simeq 10$ \citep{Mat1999,Kas2010,Mor2013}.  For explosions of red supergiants  (RSGs), one has $n\simeq 12$ \citep{Mat1999,Mor2013}. The slope of the inner region of the ejecta satisfies $\delta \simeq 0-1$. In our calculation, we adopt $\delta=0,n=10$ as fiducial values.

The dimensionless geometric factor for the ejecta mass due to the assumed density profile  distribution is a broken power law, i.e. \citep{Vin2004}
\begin{equation}
I_{\text{M}}=\frac{1}{3-\delta }x_{0}^{3}+\frac{1}{3-n}\left(
x_{0}^{n}-x_{0}^{3}\right) .
\end{equation}
The mass ratio between the inner and outer regions is
\begin{equation}
\mathcal{R}_{\rm M}\mathcal{=}\frac{3-n}{3-\delta }\left( \frac{x_{0}^{3}}{%
x_{0}^{n}-x_{0}^{3}}\right) .
\end{equation}
For $x_0=0.1$, one has $\mathcal{R}_{\rm M}=2.33$.

The dimensionless geometric factor for the kinetic energy of the ejecta is \citep{Vin2004}
\begin{equation}
I_{\text{K}}=\frac{x_{0}^{5}}{5-\delta }+\frac{1}{5-n}\left(
x_{0}^{n}-x_{0}^{5}\right) .
\end{equation}

The total optical depth of the outer region ejecta reads%
\begin{equation}
\tau _{\text{tot,out}}\left( t\right) =\tau _{\text{tot,out}}\left( 0\right) %
\left[ \frac{R_0 -R_{\min,0}}{R\left( t\right) -R_{\min}(t)}\right] ^{2},
\end{equation}
where the initial optical depth of the outer region is
\begin{equation}
\tau _{\text{tot,out}}\left( 0\right) =\kappa \rho (R_0,0) R_0 \frac{x_{0}-x_{0}^n}{n-1}.
\end{equation}

If $\tau _{\text{tot,out}}\left( 0\right) >2/3$, $R_{\rm ph}$ is located
in the outer region at early epochs.  We define a timescale $t_{\tau \text{,out}}$
when the outer part region becomes transparent ($\tau _{\text{tot,out}}\left( t\right)=2/3$), i.e.,
\begin{equation}
t_{\tau \text{,out}}= \left( \frac{R_0-R_{\min,0} }{v_{
\text{max}}-v_{\min}} \right)\left \{ \left[ \frac{3\tau _{\text{tot,out}}\left( 0\right)  }{2}\right] ^{1/2}-1 \right \}.
\end{equation}
At $t<t_{\tau \text{,out}},$ $R_{\rm ph}$ is in the outer
region, which is equivalent to $x_{\text{ph}}>x_{0}$, so that
\begin{eqnarray}
I_{\text{ph}} &=&\int_{x_{\text{ph}}}^{1}\left( x/x_{0}\right) ^{-n}dx
\notag \\
&=&\frac{x_{0}^{n}}{1-n}\left( 1-x_{\text{ph}}^{1-n}\right) \approx \frac{%
x_{0}^{n}}{n-1}x_{\text{ph}}^{1-n}.
\end{eqnarray}
We then obtain%
\begin{equation}
x_{\text{ph}}\left( t\right) =x_{0}\left[ \frac{2}{3\tau _{\text{tot,out}%
}\left( t\right) }\right] ^{\frac{1}{1-n}}.
\end{equation}

At $t>t_{\tau \text{,out}},$ the outer region becomes transparent $\left(
\tau _{\text{tot,out}}\approx 2/3\right)$. The photospheric radius $R_{\rm ph}$ would enter
the inner part region of the ejecta.
The total optical depth of the inner region reads
\begin{equation}
\tau _{\text{tot,in}}\left( t\right) =\kappa \rho (R_0,0) R_0\frac{x_{0}}{1-\delta } \left[ \frac{R_0 -R_{\min,0}}{R\left( t\right)-R_{\min}(t) }\right] ^{2}.
\end{equation}

When $t>t_{\tau \text{,out}},$  the dimensionless photospheric radius can be obtained by
\begin{equation}
x_{\text{ph}}\left( t\right) =x_{0}\left[ 1-\frac{2/3-\tau _{\text{%
tot,out}}\left( t\right) }{\tau _{\text{tot,in}}\left( t\right) }\right] .
\end{equation}

%Assumed the density profile of the ejecta is the  broken power, the photospheric radius
The $R_{\rm ph}$ and $v_{\rm ph}$ evolution are shown in Fig \ref{fig:ETA} with the black solid lines. We find that in this situation the $R_{\rm ph}$ evolution curve shares similar qualitative behaviors as the uniform density one. In particular, it shares the same decline rate after the peak. For the particular parameter set we have adopted, after $t_{\tau \text{,out}}=52.8$days, $\ x_{\text{ph}}$ occurs in the inner region, which has a slope $\delta=0$ corresponding to a constant density profile.

%Summary above discussion,

%\begin{equation}
%x_{\text{ph}}\left( t\right) =\left\{
%\begin{array}{ll}
%x_{0}\left[ \frac{2}{3\tau _{\text{tot,out}}\left( t\right) }\right] ^{\frac{%
%1}{1-n}}, & t<t_{\tau \text{,out}} \\
%x_{0}\left[ 1-\frac{\frac{2}{3}-\tau _{\text{tot,out}}\left( t\right) }{\tau
%_{\text{tot,in}}\left( t\right) }\right] , & t_{\tau \text{,out}}\leq t\leq
%t_{\tau }%
%\end{array}%
%\right.
%\end{equation}

\subsection{CASE III:  exponential density profile}
We now consider the density profile in the form of
\begin{equation}
\eta(x)= \exp(-ax),
\end{equation}
where $a$ is a small positive value, with $a=1.72$ representing the Pacyz\'nski red supergiant envelope \citep{Arn1980}.

In this case, the dimensionless geometric factors for the ejecta mass and the kinetic energy are as follows:
\begin{equation}
I_{\rm M}= \frac{2-(a^2+2a+2)e^{-a}}{a^3},
\end{equation}
and
\begin{equation}
I_{\rm K} =\frac{24+\{-24-a[24+a(4+a)] \}e^{-a}}{a^5}.
\end{equation}

Similar to the above analysis, we can obtain the dimensionless photospheric radius as
\begin{equation}
x_{\rm ph}(t)= - \frac{1}{a} \ln \left[ \frac{2 (1- e^{-a})}{3 \tau_{\rm tot}(t)} + e^{-a} \ \right],
\end{equation}
where the total optical depth  is
\begin{equation}
\tau _{\text{tot}}\left( t\right) =\kappa \rho (R_0,0) R_0 \frac{1-e^{-a}}{a}  \left[ \frac{R_0 -R_{\min,0}}{R\left( t\right)-R_{\min}(t) }\right] ^{2}.
\end{equation}

%For the exponential density profile, the photospheric radius
The $R_{\rm ph}$ evolution for this case is shown in Fig \ref{fig:ETA} as the green solid curve.  Since the density gradient $d\eta/dx$ is larger than that of the uniform density profile, the $R_{\rm ph}$ decline rate is much slower.

\subsection{CASE IV:  density increases with radius}
In the three cases mentioned above, the density profile of the ejecta is either a constant or decreasing with the radius. It is interesting to investigate the opposite case, i.e. the density increases with radius so that there is a positive density gradient, even though it is difficult to realize such a density profile in SN explosions.

We assume the density profile as a power law, i.e.,
\begin{equation}
\eta(x)=x^{m},
\end{equation}
where the power law index $m$ is assumed to be a positive value to allow density increasing with radius. The uniform density profile corresponds to $m=0$.

In this situation, the photospheric radius reads
\begin{equation}
R_{\text{ph}}\left( t\right) =R\left( t\right) \left( 1-\frac{2}{3\tau _{%
\text{tot}}\left( t\right) }\right) ^{\frac{1}{m+1}},
\end{equation}%
where%
\begin{equation}
\tau _{\text{tot}}\left( t\right) =\frac{\kappa \rho (R_0,0)
R_0 }{1+m} \left[ \frac{R_0 -R_{\min,0}}{R\left( t\right)-R_{\min}(t) }\right] ^{2}.
\end{equation}
The  time derivative of the photosphere is
\begin{eqnarray}
\frac{dR_{\text{ph}}}{dt} &=&(v_{\text{max}}-v_{\text{min}})\left[ 1-\frac{2}{3\tau _{\text{tot}}\left( t\right) }\right]
^{\frac{1}{m+1}}  \notag \\
&&-\frac{4(v_{\text{max}}-v_{\text{min}})}{3\left( 1+m\right) }\frac{1}{\tau _{\text{tot}%
}\left( t\right) }\left[ 1-\frac{2}{3\tau _{\text{tot}}\left( t\right) }%
\right] ^{\frac{-m}{m+1}}.
\end{eqnarray}
We find that when $\tau _{\text{tot}}=\tau _{\text{tr}}=2\left(
m+3\right) /3\left( m+1\right)$, the photospheric radius reaches its peak.

We adopt $m=2$, the $R_{\rm ph}$ and $v_{\rm ph}$ evolution  are shown in Fig \ref{fig:ETA} as the blue solid lines. Compared with the three cases mentioned in previous subsections, the photospheric radius in this case decreases very rapidly after the peak due to the rapid decrease of density as the photosphere recedes in the ejecta.
% recession of$\ x_{\text{ph}}$, the density is getting smaller and smaller.

\section{Conclusions and Discussion }
\label{sec:dis}

\begin{figure}[tbph]
\begin{center}
\includegraphics[width=0.45\textwidth,angle=0]{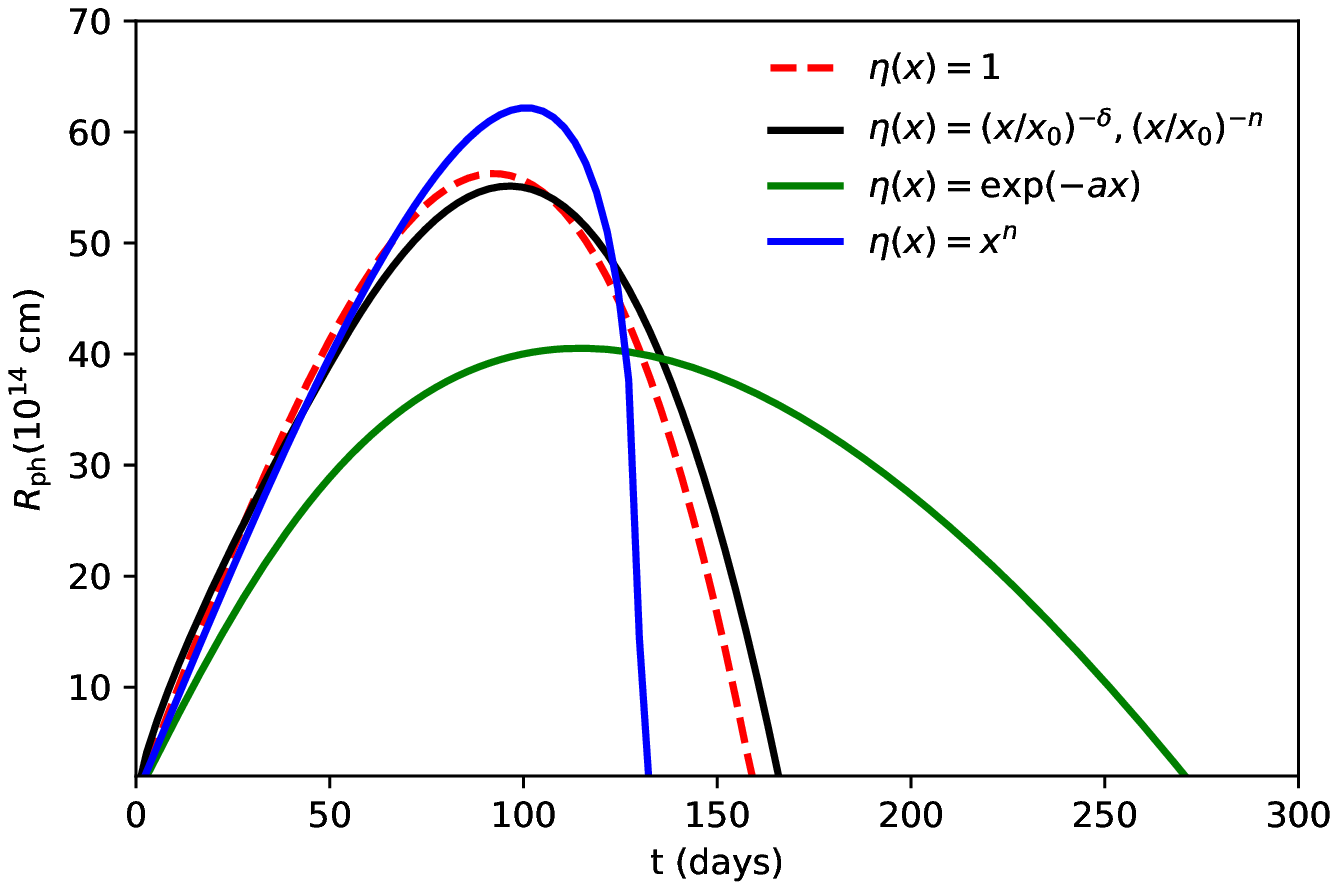}
\includegraphics[width=0.45\textwidth,angle=0]{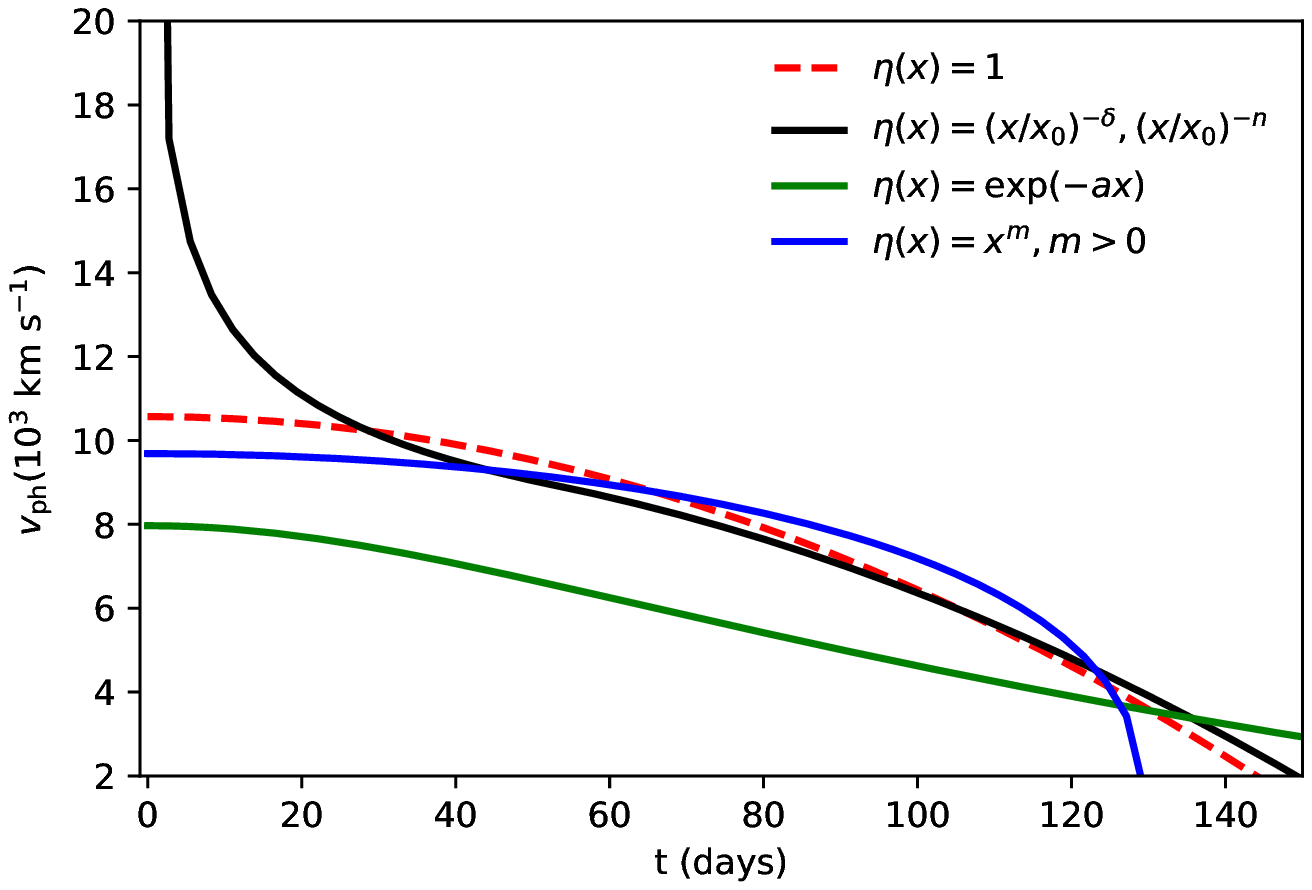}
\end{center}
\caption{ \textbf{ The photospheric radius (left panel) and velocity (right panel) evolution with the different density profiles for the following choice of parameters: $M_{\rm ej}=3.0M_{\odot}$; $E_{\rm K}=2\times 10^{51}$ erg; $R_0=10^{13}$cm; and $\kappa = 0.1$ cm$^{2}$ g$^{-1}$.}}
\label{fig:ETA}
\end{figure}

\begin{figure}[tbph]
\begin{center}
\includegraphics[width=0.85\textwidth,angle=0]{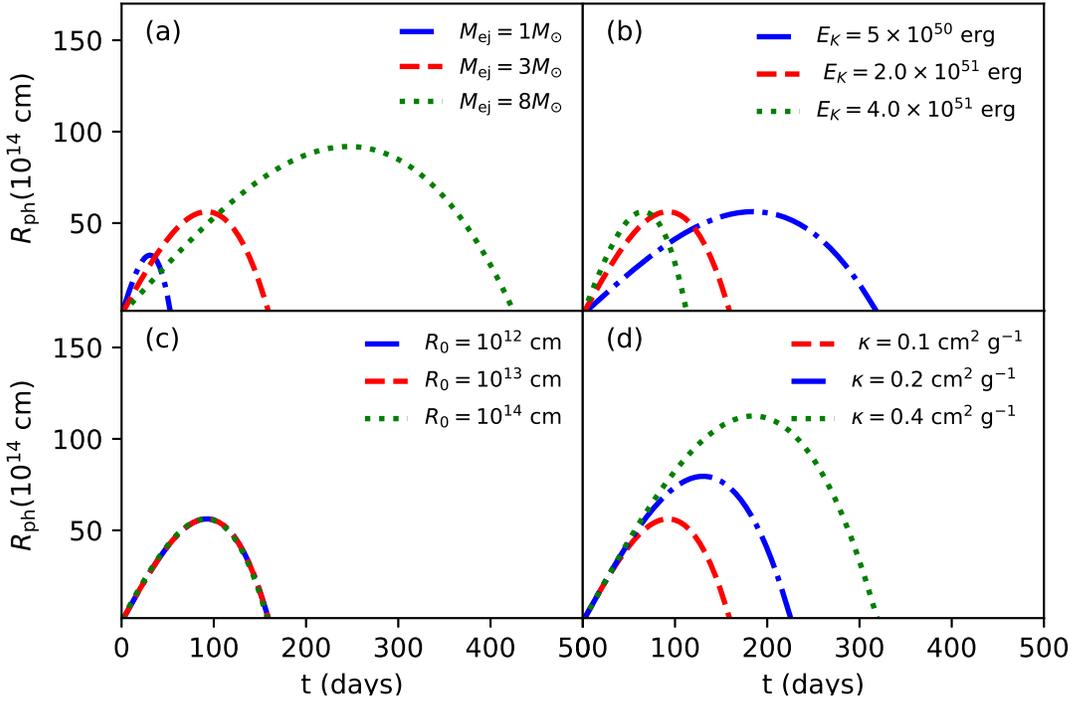}
\end{center}
\caption{Effects of the varying ejecta mass $M_{\rm ej}$ (panel a), kinetic energy $E_{\rm K}$  (panel b),  initial radius $R_0$ (panel c), and  opacity (panel d) of the ejecta. The uniform density profile $\eta(x)=1$ is adopted. The fiducial parameters are (plotted with red dashed line): $M_{\rm ej}=3.0M_{\odot}$; $E_{\rm K}=2\times 10^{51}$ erg; $R_0=10^{13}$cm; and $\kappa = 0.1$ cm$^{2}$ g$^{-1}$. }
\label{fig:MRE}
\end{figure}

We have investigated a general model of homologous expansion with an
arbitrary density distribution profile and the evolution of the photospheric radius. We discover a generic behavior, i.e. $R_{\rm ph}$ always rises at early epochs  and falls at late epochs. As shown in Fig \ref{fig:ETA}, different density profiles affect the shape of the $R_{\rm ph}$ evolution curves, especially the rate of decline after the peak. However, the general qualitative behavior remains the same. Investigating how various parameters might affect the $R_{\rm ph}$ evolution curve (Fig \ref{fig:MRE}), we find
 %shows the behavior of the photospheric radius  for different input parameters. We find
 that the initial radius has a negligible effect, while ejecta mass, kinetic energy, and opacity all influence the maximum $R_{\rm ph}$ and the time to reach the peak.
 %And given a specific density profile, the rate of decline after the peak is almost independent on the input parameters.

Our treatment assumed a constant opacity.
In general, the opacity is a function of density, temperature, and composition of the ejecta.  It is essentially a constant when Thomson scattering  dominates the opacity. If the local temperature of the ejecta drops below the recombination temperature $T_{\rm rec}$,  the ejecta is  mostly neutral, in which case the opacity is almost zero.  Taking the recombination effect into account, the ejecta becomes transparent in a shorter time scale \citep{Arn1989}. Considering the effect of the recombination would introduce additional complications of $R_{\rm ph}$ evolution, which is not investigated in this paper.
% the general behavior of the photospheric radius evolution.

So far we have ignored emission from the outer layers in the nebular phase, so the above theory is applied to the case when the emission from the nebular phase does not outshine the emission from the photosphere. At late epochs of an explosion, such an assumption is no longer valid. On the other hand, the observed spectrum would deviate from blackbody since the emission is optically thin. For an ideal observational campaign with wide-frequency-band observations, such a phase can be in principle identified. In practice, photometric observations in several different colors may not be able to tell the difference, so that an ``effective'' $\tilde R_{\rm ph}(t)$ is derived based on the observed $L_{\rm bol}(t)$ and $T_{\rm eff}(t)$, which include the contributions from both the true photosphere and gas above. This is not the true photospheric radius, which decays slower than the true $R_{\rm ph}(t)$. This explains the shallow $R_{\rm ph}$ decay in many transients as revealed by observations \citep[e.g.][]{Nic2016,Das2018}.

Some SNe, especially SNe IIn, show evidence of interaction between the SN ejecta and the circumstellar medium (CSM) around the progenitor. In this case, since the velocity of the outer layers of the SN ejecta is much higher  than the velocity of the CSM, one may assume that the ejecta interacts with a relatively stationary CSM. The photospheric radius is located in the CSM rather than in the SN ejecta. Photons diffuse through an optically thick CSM with a fixed photosphere \citep{Cha2012,Cha2013}. Therefore, in this situation, one has $T(t) \propto  L_{\rm bol}^{1/4}(t)$. The difference in the $R_{\rm ph}$ evolution behaviors between the interacting model and the homologous explosion model discussed here can be used to diagnose the physical origin of an observed SN event.

%Recently, an unusual, luminous, fast evolving transient AT 2018cow was discovered in the local universe. \citep{Pre2018,Per2018,Kui2018}. The nature of  AT 2018cow is still unclear.

\begin{figure}[tbph]
\begin{center}
\includegraphics[width=0.85\textwidth,angle=0]{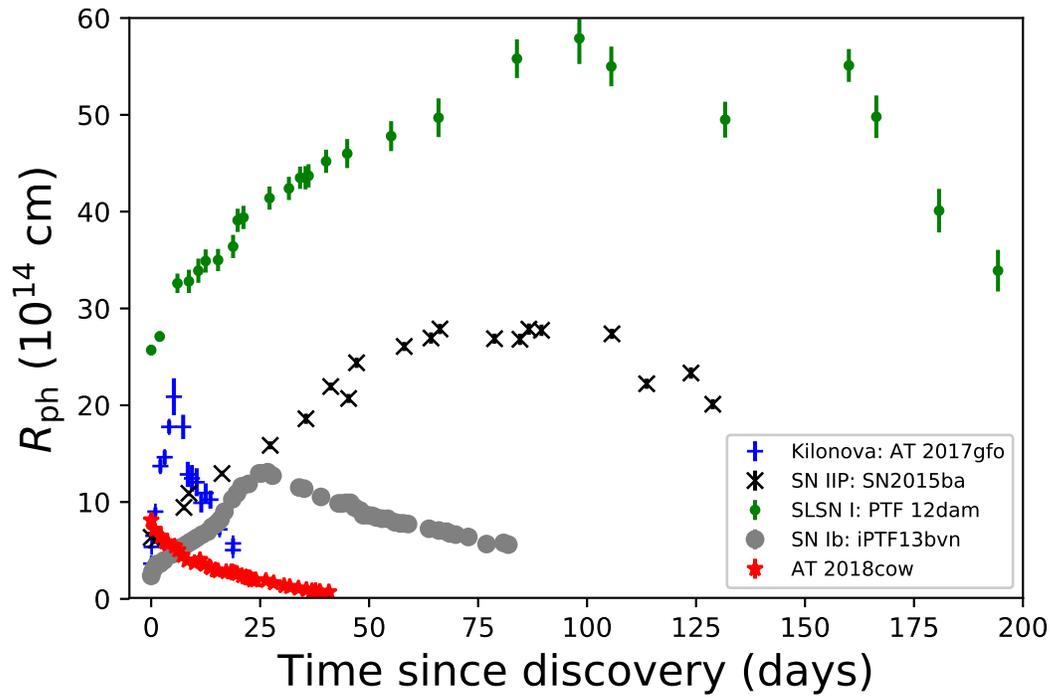}
\end{center}
\caption{  The photospheric radius evolution of various optical transients: Type I superluminous SN PTF12dam, Type IIP SN 2015ba, Type Ib SN  iPTF13bvn, the kilonova AT 2017gfo associated with GW170817/GRB 170817A, as well as the peculiar event AT 2018cow that is likely not from an explosion. }
\label{fig:obse}
\end{figure}

In Fig \ref{fig:obse}, we collect a sample of explosions whose $R_{\rm ph}$ evolution is well observed. One can see that the general rising/falling behavior of $R_{\rm ph}$ predicted in our theory is found in different types of SNe, including superlumious supernova PTF 12dam \citep{Vre2017}, Type IIP supernova SN 2015ba with a long plateau \citep{Das2018}, and Type Ib iPTF13bvn\citep{Fre2016}.
The photospheric radius of the ``kilonova'' transient AT2017gfo associated with GW170817 also exhibited such a rising/falling behavior \citep{Dro2017}. The widths of the $R_{\rm ph}$ peaks depend on the physical parameters of the explosions (e.g., $M_{\rm ej}$, $E_{\rm K}$, and $\kappa$), but the general evolution behavior is similar.

The special event AT 2018cow shows a peculiar behavior of steady decline of $R_{\rm ph}$ as a function of time \citep{Per2018,Kui2018}, see Fig \ref{fig:obse}.According to the theory discussed in this paper, this behavior makes AT 2018cow essentially impossible to be a supernova. Observationally, $R_{\rm ph}$ decays from the very beginning, and no rising $R_{\rm ph}$ was detected. In order to interpret the source as a SN, the ejecta mass should be very small, e.g. $\sim 0.05 M_{\odot}$, in order to make a very rapid rise to satisfy the observational constraint \citep{Pre2018}. For such a small mass, the ejecta would become transparent in a very short period of time, e.g. $t_{\tau} =3.2$d for uniform density distribution. However, observationally, the photospheric radius of AT 2018cow continually decreases over a much longer period of time ($>30$ d) since the first detection. If the emission is from the nebula phase, the effective photospheric radius $\tilde R_{\rm ph}$ would display an increasing trend due to the expansion of the ejecta, in contrary to the observations. Our results support its interpretation within the framework of a tidal disruption event \citep{Per2018,Kui2018}.

%\textbf{Observations usually reveal supernova whose increase of photospheric radius is faster than its decline \citep{Nic2016,Das2018}. So the CASE IV, the density increases with radius is unlikely to happen in the SN ejecta. When $t>t_{\tau}$  the ejecta becomes mostly optically thin. Although the nebular emission is not blackbody, it is largely thermal emission. At late-time, the observed photospheric radius is definitely not the blackbody radius we calculated in our paper. }

If the ejecta is a radiation-dominated gas, a strictly adiabatic cooling solution would give $T \propto   R(t)^{-1}$. According to \cite{Arn1980,Arn1982}, the temperature distribution within the ejecta can be described as
\begin{equation}
T^{4}\left( r,t\right) =T^{4}(R_0,0) \Psi \left( x\right) \phi
\left( t\right)  \left[ \frac{R_0 -R_{\min,0}}{R\left( t\right)-R_{\min}(t) }\right] ^{4}.
\label{Tevo}
\end{equation}
where $\phi (t)$ is the temporal part solution of energy conservation equation of the expanding ejecta,  while $[(R(0)-R_{\min,0})/(R(t)-R_{\rm min}(t))]^4$ describes the adiabatic cooling of the ejecta. The spatial part of the solution $\Psi (x)$ depends on the density profile $\eta(x)$. Observationally, the evolution of $R_{\rm ph}$ can be directly used to constrain the density profile of the ejecta if the contamination from the gas above the photosphere is insignificant or can be removed. The observed photospheric temperature as a function of time, when coupled with the inferred density profile as well as the adiabatic evolution law in Eq.(\ref{Tevo}), can be used to directly diagnose the temperature structure of the ejecta. Direct confrontations of our theory with detailed observational data of diverse explosion events will be carried out in future work.

% Compared the observation photosphere  with  the theoretical photospheric  radius evolution could help us to diagnose the density profile of supernovae ejecta. If we know the density distribution of the ejecta, we could infer the temperature distribution from observation.

\acknowledgments

 We thank Simon Prentice, Danial Perley for helpful information,
% for sharing the bolometric light curve data of AT 2018cow and helpful discussion.   We thank
 Shanqin Wang, Shaoze Li, Tony Piro for discussion, and the referee for helpful suggestions.
This work was supported by the National Basic Research
Program (``973" Program) of China (grant No. 2014CB845800) and the National
Natural Science Foundation of China (grants No. 11573014 and 11833003). This work was also
supported by the National Program on Key Research and Development Project of
China (grants no. 2017YFA0402600 and 2016YFA0400801). L.D.L.
%and S. Z. Li are
is supported by a scholarship from
the China Scholarship Council (No. 201706190127)
%and No. 201706770050)
to conduct research at the University of Nevada, Las Vegas (UNLV).

%
%\clearpage
%
%
%\begin{table}
%%\begin{sidewaystable}[tbph]
%\caption{Fitting parameters for AT 2018cow with the ejecta-CSM interaction model}
%\begin{center}
%\begin{tabular}{cccccc}
%\hline \hline
%$E_{\rm SN}$ & $M_{\rm ej}$ & $M_{\rm CSM}$ & $\rho_{\rm CSM,in}$ \tablenotemark{a}& $R_{\rm CSM,in}$ & $\epsilon$ \\
%$10^{51}$ erg & $M_{\odot}$ &  $M_{\odot}$ & $10^{-12}$ g cm$^{-3}$& $10^{14}$ cm &  \\ \hline
%$1.7^{+0.21}_{-0.27}$& $2.84^{+0.92}_{-0.66}$& $0.17^{+0.01}_{-0.01}$& $1.47^{+0.15}_{-0.11}$&$2.73^{+0.33}_{-0.29}$&$0.61^{+0.10}_{-0.14}$\\
%\hline \hline
%\end{tabular}%
%\end{center}
%\par
%a. $\rho_{\text{CSM,in}}$ is the density of CSM at the inner radius $R_{\rm CSM,in}$.
%\label{tbl:fitting par}
%%\end{sidewaystable}
%
%\end{table}

\clearpage

\end{document}